\documentclass{IEEEtran}
\usepackage{cite}
\usepackage{amsmath,amssymb,amsfonts}
\usepackage{graphicx}
\usepackage{textcomp,nicefrac}
\usepackage{url}
\usepackage[frozencache,cachedir=.]{minted}

\usepackage{listings}
\usepackage{tikz}
\usepackage[most]{tcolorbox}

\usepackage{pgfplots}
\usetikzlibrary{patterns}

\usepackage{siunitx} 

\usepackage{subcaption}

\def\BibTeX{{\rm B\kern-.05em{\sc i\kern-.025em b}\kern-.08em
T\kern-.1667em\lower.7ex\hbox{E}\kern-.125emX}}
\begin{document}
\title{BORA: A Personalized Data Display for Large-scale Experiments}
\author{Nicholas Tan Jerome, Suren Chilingaryan, Timo Dritschler and Andreas Kopmann
\thanks{Manuscript received June XX, 20XX; revised November XX, 20XX; accepted
December XX, 20XX. Date of publication January XX, 20XX; date of current
version July XX, 20XX.}
\thanks{N. Tan Jerome, S. Chilingaryan, T. Dritschler, and A. Kopmann are with the Institute
for Data Processing and Electronics, Karlsruhe Institute of Technology,
76131 Karlsruhe, Germany (e-mail: nicholas.tanjerome@kit.edu).}
}

\maketitle

\begin{abstract}
Given the rapid improvement of the detectors at high-energy physics experiments, the need for real-time data monitoring systems has become imperative. The significance of these systems lies in their ability to display experiment status, steer software and hardware instrumentation, and provide alarms, thus enabling researchers to manage their experiments better. However, researchers typically build most data monitoring systems as standalone in-house solutions that cannot be reused for other experiments or future upgrades. We present BORA (personalized collaBORAtive data display), a lightweight browser-based monitoring system that supports diverse protocols and is built specifically for customizable visualization of complex data, which we standardize via video streaming. We show how absolute positioning layout and visual overlay background can address the diverse data display design requirements. Using the client-server architecture, we enable support for diverse communication protocols, with the server component responsible for parsing the incoming data. We integrate the Jupyter Notebook as part of our ecosystem to address the limitations of the web-based framework, providing a foundation to leverage scripting capabilities and integrate popular AI frameworks. Since video streaming is a core component of our framework, we evaluate viable approaches to streaming protocols like HLS, WebRTC, and MPEG-Websocket. The study explores the implications for our use case, highlighting its potential to transform data visualization and decision-making processes. 
\end{abstract}


\section{Introduction}
\label{sec:introduction}

\IEEEPARstart{R}{eal-time} data monitoring plays a crucial role in modern research to ensure the integrity and quality of experiments. Through constant monitoring and prompt detection, issues like equipment malfunctions, excessive signal noise, and data corruption can be detected and addressed at runtime. This is highlighted by experiments like the discovery of the Higgs boson~\cite{cms2022portrait}, which required long measurement and observation times due to the low statistical likelihood of relevant events occurring. Such experiments and similar endeavors requiring large statistic sample sizes~\cite{krizan2022belle,cepeda2019higgs,houdy_2020} lead to the generation of petabytes or even exabytes of data, thereby presenting challenges to their respective data monitoring systems.

With the increase in data rates from modern detector hardware, the traditional approach of "store-first-analyze-later" is no longer feasible. In particular, scientists need to adjust instrumentation parameters and make real-time decisions regarding the recorded data's relevance. Ideally, we want to observe and interact even with large data streams. In response, experiments adopt distributed computing technologies to manage complex systems with multiple components, with the data monitoring component built as a standalone view-only user displays~\cite{chen2019scientific}.

Most existing monitoring systems use static views to visualize the system status, incorporating knowledge about specific instrumentation~\cite{allan2012omero,taghizadeh2020sciserver}. Although there are established protocols for exchanging parameters and slow-control information, there is currently no standard approach for presenting detector data to the operator, thereby separating data acquisition from the monitoring system.

In this paper, we present techniques to facilitate data access for web displays, following the principle that emphasizes small upfront investment (development effort) and enables rapid deployment. We realize our techniques in BORA (personalize colla\underline{bora}tive data display), a browser-based monitoring system that supports diverse protocols and is specifically for customizable visualization of complex data, which we standardize as
video streaming. Additionally, we use Jupyter notebooks to interact with data and make dynamic changes to web views. These notebooks allow users to manipulate data and update web view settings through scripting. With this feature, the operator can script functions that are not natively supported by the monitoring system.

\begin{figure*}
  \begin{subfigure}{0.5\textwidth}
    \includegraphics[width=\linewidth]{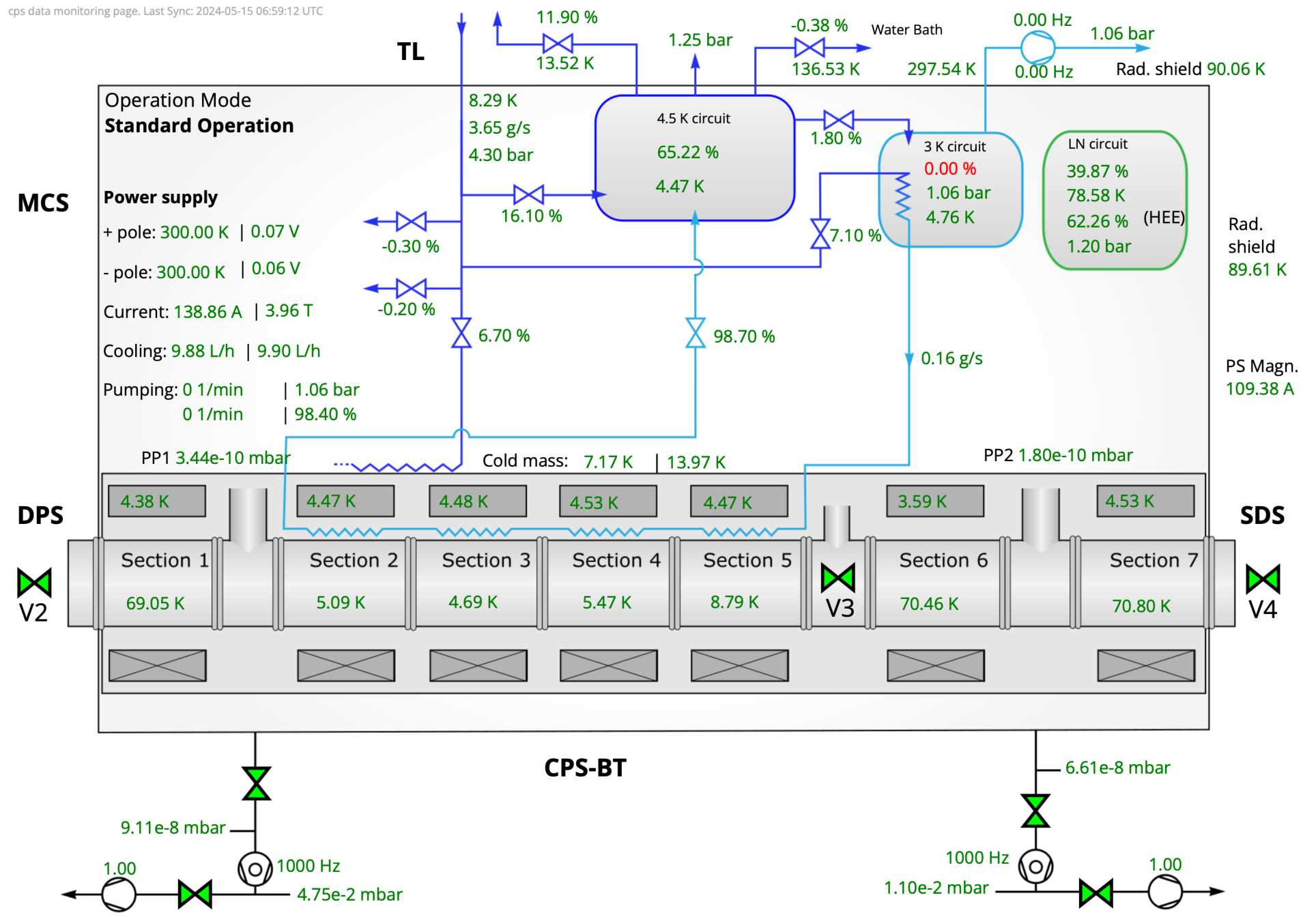}
    \caption{} \label{fig:1a}
  \end{subfigure}%
  \hspace*{\fill}   
  \begin{subfigure}{0.5\textwidth}
    \includegraphics[width=\linewidth]{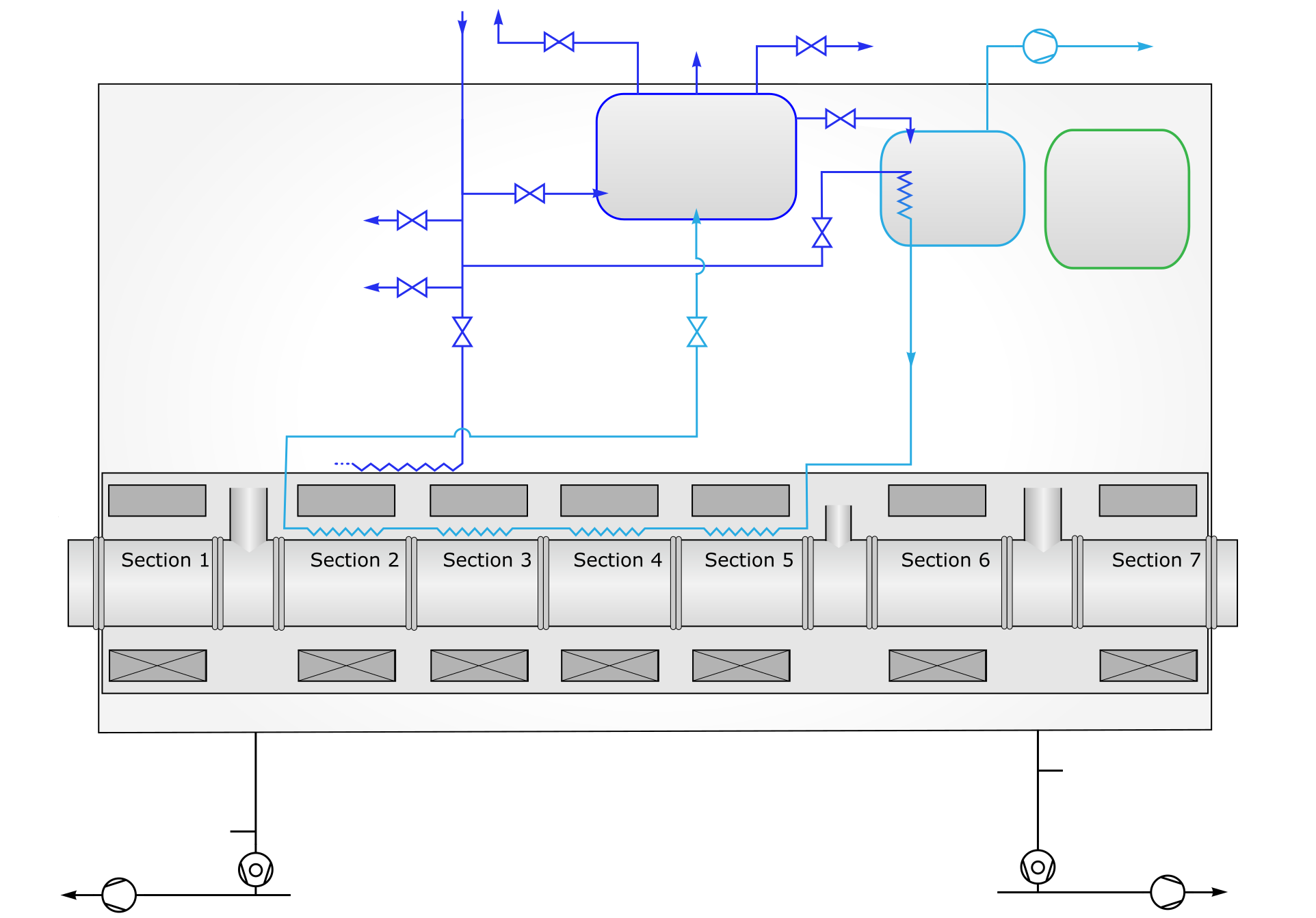}
    \caption{} \label{fig:1b}
  \end{subfigure}%
\caption{(a) A BORA status monitoring display for the Cryogenic Pumping Section segment of the KATRIN experiment and (b) the corresponding background image.}\label{fig:cps_display}
\end{figure*}

\section{Related Work}

Many real-time monitoring systems have been developed for high-energy physics experiments to ensure data quality, real-time analysis, and resource optimization. These systems constantly evaluate data integrity, facilitating prompt detection of anomalies and adaptation of data collection strategies. While real-time monitoring systems are integral to scientific experiments, there is a lack of consensus on their design, resulting in the need for individualized in-house solutions for each experiment. In this section, we explore the approaches taken by other experiments in implementing their data monitoring systems.

The management of large-scale data in high-energy physics is exemplified by the Large Hadron Collider (LHC)~\cite{barisits2019rucio}, which deploys the Data Quality Monitoring (DQM) software as a core component of the CMS experiment. The DQM enables real-time detector monitoring, provides prompt offline feedback for data quality analysis and certification, validates all reconstruction software, and validates Monte Carlo productions. Data status is shown in multiple web views (DQMGUI). Each DQMGUI comprises a web server hosting a static web-based user interface~\cite{azzolini2019data}. 

The Belle II experiment~\cite{konno2015slow} uses the DQM system to create and update histograms by analyzing online event samples and to visualize the histograms on live browser GUIs and update them at the interval of ten seconds. The analytical part of the DQM system is based on the BASF2~\cite{moll2011software}, implemented as ROOT~\cite{antcheva2009root} based C++ applications and operated from Python scripts, while the visualization part is based on C++ and Java.

The DAQ monitoring system for the vertex detector prototype~\cite{li2024beam} at Circular Electron Positron Collider (CEPC)~\cite{cepc2018cepc} uses a modular design approach to monitor experiment status, configure file generation, manage message logging, and display online hit maps.

The LEPS2 experiment~\cite{ryu2020current} integrates its monitoring system using the middleware-based architecture, where the client GUI is served by the DAQ Operator through an HTTP server. The client GUI feeds the data from the DAQ-middleware to display histogram plots and experiment data.

The Jiangmen Underground Neutrino Observatory (JUNO)~\cite{abusleme2021juno} implemented three standalone web displays for instrumentation data monitoring, a control panel, and an alarm system. Utilizing client-server and publish-subscribe technologies, the data monitoring system effectively moves data from the EPICS system to a MySQL database, which is subsequently pushed to the web-based clients~\cite{mei2016framework}. The JUNO device control is based on the Vue3 framework written in TypeScript, using HTTP and Websocket for data exchange with the server. The JUNO alarm is based on the Kafka framework.

This highlights how disjunct and diverse the technologies and standards of contemporary monitoring systems have become. However, regardless of the different technologies or frameworks in these monitoring systems, they share similarities wherein all of these systems have status, control, and alarm displays. Also, these displays are deployed as modular components fitted to a distributed computing ecosystem. The modular approach is necessary to maintain a complex system.

While these in-house solutions can have functionalities tailored precisely to the needs of their experiment, newer monitoring systems look into open-source solutions to provide status, control, and alarm displays. The Deep underground neutrino experiment (DUNE)~\cite{falcone2022deep} and ATLAS~\cite{aad2008atlas} experiments have chosen Grafana~\cite{chakraborty2021grafana} as their monitoring systems; the monitoring system most similar to BORA.

Grafana provides a flexible and customizable platform for visualizing data from diverse sources, offering interactive dashboards tailored to specific needs. Its unified monitoring capabilities aggregate data from multiple sources into a single interface, enabling comprehensive monitoring and visualization of system performance. However, the ATLAS computing team reported a limited set of visualizations, which does not complement the display requirements of their experiments~\cite{beermann2020implementation}. 

Existing data monitoring systems based on static views present system status, enable limited device control and provide alarm features. These systems necessitate knowledge about particular instrumentation, making them a one-time-off in-house solution. Our work extends and integrates these approaches. Our system assumes a client-server architecture, wherein the server is tasked with data reduction processes that require significant computational resources. Our framework arranges data to match the user's envisioned displays, unhampered by the visual limitations imposed by the framework.

\section{Design Concepts}

Our approach to designing an effective experiment monitoring system follows an overarching principle that scientists should not be limited by the capability of the software framework. Here, we present our rationale for designing a multi-purpose data quality monitoring system and discuss our idea of encoding large-scale data into video streams. Given the uniqueness of each experiment, we include Jupyter notebooks~\cite{randles2017using} that can provide user-space data manipulation and extend BORA setups with scripting capabilities. This empowers the operator to script anything that BORA does not support natively. The BORA ecosystem also provides the foundation to include artificial intelligence, such as utilizing large language models with Jupyter notebooks~\cite{randles2017using,brown2020language}, and performing machine learning techniques in data correlation and forecasting~\cite{yang2012improved,gao2022reinforcement,yu2022pseudo}.

\subsection{Absolute Positioning}

In web design, there are widely accepted guidelines on how to present information effectively without overwhelming users. This consensus of best practices in web design leads to a fixed layout approach, where visual graphs are represented as boxes stacked together in a dynamic layout. This is the path that Grafana took, which requires a high upfront investment (development effort). Users must familiarize themselves with the framework and are restricted to its provided functionalities. We adopt a contrasting design philosophy that aims to minimize the initial upfront investment, thereby reducing the development effort. Therefore, we simplify the complexity by sacrificing certain functionalities, like the stacked dynamic layout approach.

We adopt absolute positioning for the front-end layout to emphasize simplicity in the design. A fixed x-y coordinate determines the placement of each widget. However, this simplicity also poses the disadvantage of less adaptive screen layouts. Adjusting the display to a specific display resolution is contingent upon using the native browser zoom capability. 

Another notable design decision is incorporating a background image to address the difficulties of maintaining a no-code user interface design page. Rather than displaying visually intricate elements to users, the framework employs a background image as an overlay in response to user input. Including this feature enables the BORA framework to adapt to various experiments with different specifications. The BORA display depicted in Figure~\ref{fig:cps_display} uses a background image and employs absolute positioning to place each visual widget according to x-y coordinates.

\subsection{Jupyter Notebook for Data Interaction and Web Views}

Interaction is essential to control the extraction of information from the data displays. Still, the continuous development of an experiment demands constant changes and updates, thus leading to many customized solutions. While modular standalone architecture may cope with continuous updates, we propose the integration of Jupyter notebooks as part of the web-based display's ecosystem. The usage of notebooks has gained increasing attention in the scientific community, where scientists mainly use notebooks as "scratchpads," from which code is later extracted to scripts. These notebooks can be shared, simplifying the transfer and exchange of results, algorithms, and code.~\cite{kery2018story, pimentel2019large}.

\begin{figure}[t]
      \includegraphics[width=0.48\textwidth]{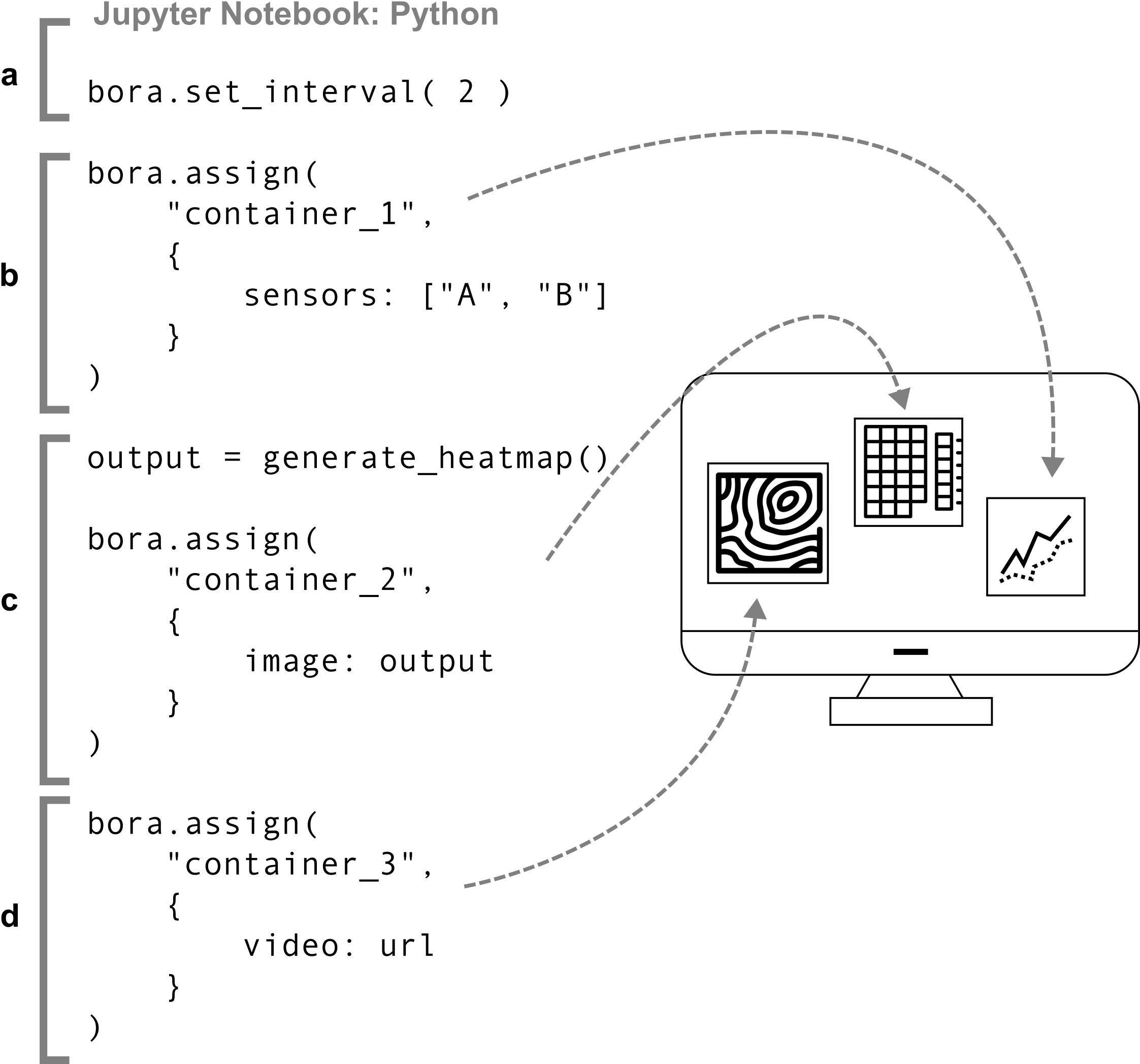}
  \caption{Jupyter integration with BORA. (a) Define the data polling interval of BORA (2 seconds). (b) Add a list of sensors to the BORA widget with the identifier "container\_1". (c) Generate a heatmap using a user-defined function, which pulls the data from a Redis database. Then, attach the output image to the BORA widget with the identifier "container\_2". (d) Add a video URL to the BORA widget with the identifier "container\_3".}
  \label{fig:bora_jupyter}
\end{figure}

Users have traditionally been able to configure the data display parameters through setup files, for example, in the XML format~\cite{ryu2020current}. A limitation is associated with this, as updating the setup parameters necessitates a system restart, which is inconvenient when conducting an active experiment. Instead, we suggest employing Jupyter Notebooks to provide supplementary techniques for modifying the display settings at runtime and interactively updating the data visualization. Additionally, users can manipulate data and assign the resulting output of the customized function to a visual block. Figure~\ref{fig:bora_jupyter} shows the Jupyter notebook integration that updates the BORA settings and corresponding widgets. We follow the approach shown by Clarke et al.~\cite{clarke2021appyters}, where we turn Jupyter notebooks into fully functional web-based data inspection applications. By inspecting the data in our database, we can perform correlation, forecasting, pattern detection, and much more. We can also create reusable workflows, including applications to build customized machine learning pipelines, analyze recent data, and produce publishable figures. 

\subsection{Encode Large Data into Video Streams}

We provide the option to encode large data within video streams, a highly relevant capability in high-speed imaging applications. High-speed imaging plays a critical role in a wide range of scientific and industrial fields because of its ability to capture events that are otherwise impossible to observe with the naked eye or standard imaging techniques~\cite{suzuki2017single,velten2012recovering}. As commercial high frame rate cameras are becoming affordable, more and more researchers seize the opportunity to explore new knowledge using these devices. However, these cameras deliver high frame rates that are challenging to make sense of without appropriate tools. Typically, commercial CCD (Charge-Coupled Device) and CMOS (Complementary Metal-Oxide-Semiconductor) systems could deliver a frame rate that exceeds \SI{10}{\kilo fps} (frames-per-second)~\cite{versluis2013high}.

These high-speed cameras can continuously stream data over 10~GB per second, complicating data storage. There is a lack of studies and solutions regarding managing such large data streams in process monitoring and quality control. Many researchers have resolved to the conventional data warehouse store-first-analyze-later approach, where high data rates can quickly exhaust memory bandwidth and become performance bottlenecks~\cite{kraska2013finding}.

To deal with such a high frame rate, the system can process the incoming frames in high-performance computation pipelines such as gStreamer~\cite{taymans2013gstreamer,angsuchotmetee2018pipelining}. A practical solution is to implement a shared memory ring buffer framework to handle the high volume of incoming frames. This framework can then distribute the frames to other subsystems for additional processing~\cite{ingles2017c++}. The client system could then visualize the processed data at a more user-friendly frame rate, such as 60 Hz~\cite{jiang2022user}. The computation in the pipeline mainly focuses on reducing data and extracting features. We propose to use video streaming to integrate various types of complex data and look to enable supporting video-streaming protocols in our framework.

Our primary interest lies in existing real-time video streaming protocols that web browsers can support. We explore three methods for real-time video streaming that prioritize low-latency communication: HLS (HTTP Live Streaming)~\cite{pantos2017http}, MPEG-Websocket~\cite{wu2017low}, and WebRTC (Web Real-Time Communication)~\cite{nurminen2013p2p}. HLS is a streaming protocol that splits media content into smaller segments and sends them through HTTP. The HLS streaming protocol is receiving significant attention in the scientific community for its fast streaming capabilities~\cite{durak2020evaluating,zhang2021performance}. HLS enables dynamic quality level switching and adaptive bitrate streaming based on network conditions. The output file, which is an M3U8 playlist, is used to supply clients with metadata and segment URLs. The second approach involves integrating WebSocket with MPEG. WebSocket is valued for its ability to create low-latency video streaming systems. We leverage WebSocket channels for transmitting MPEG-encoded content, which will then be decoded in the client browser~\cite{wu2017low}. This combination is frequently used in applications requiring rapid, bidirectional communication, such as live streaming, video conferencing, and interactive web experiences. WebRTC offers a method of directly connecting the streaming source and the client browser. To establish a low-latency connection, it uses WebSocket for signaling and gathering information required to tunnel from source to destination~\cite{blum2021webrtc}.

\section{Evaluation of Video Streaming Approaches}

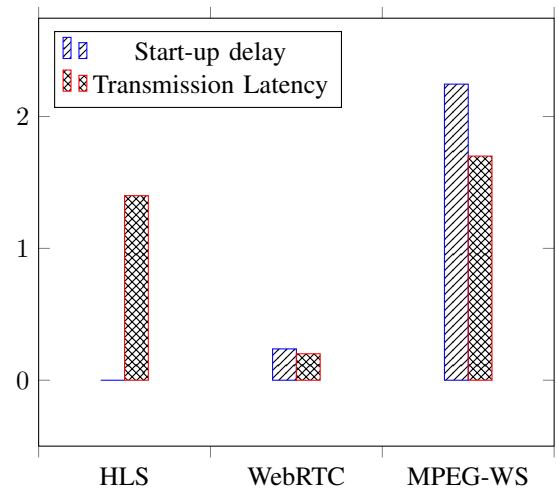
\begin{figure}[t]
  \centering
\begin{tikzpicture}
\begin{axis}[
ymin=0,
legend pos=north west,
enlargelimits={abs=0.5},
ybar=0pt,
bar width=9,
xtick={0.5,1.5,...,5.5},
xticklabels={HLS,WebRTC,MPEG-WS},
x tick label as interval
]
\addplot+[
    pattern=north east lines,
    error bars/.cd,
    y dir=both,y explicit
] coordinates {
    (1,0)
    (2,0.237509)
    (3,2.245823)};

\addplot+[
    pattern=crosshatch,
    error bars/.cd,
    y dir=both,y explicit
] coordinates {
    (1,1.4)
    (2,0.200556)
    (3,1.700054)};
\legend{Start-up delay, Transmission Latency}
\end{axis}
\end{tikzpicture}
\caption{Comparison of the start-up delay and transmission latency between the HLS approach, the MPEG-Websocket approach, and the WebRTC approach.}
  \label{fig:vis_res}
\end{figure}

Since BORA is built to support video streaming, we evaluate different video streaming technologies to better select the right technology. To this end, we created a C++ prototype RTSP source using the GST template library. The prototype generates a video stream showing a sphere circling within a space. The RTSP source is accessible through an endpoint and can be utilized for various streaming techniques. 


For the case of HLS, we use FFMPEG to transcode the RTSP protocol to a M3U8 playlist. The following command is used to generate the playlist file:

\begin{minted}{bash}
ffmpeg -i "rtsp://127.0.0.1:8554/test" \
       -hls_time 3 \ 
       -hls_wrap 10 \
       "stream/streaming.m3u8"
\end{minted}

\noindent where \mintinline{bash}{rtsp://127.0.0.1:8554/test} is the endpoint of the RTSP source, and \mintinline{bash}{stream/streaming.m3u8} is the playlist output file. This command doesn't specify the transport protocol (\mintinline{bash}{-rtsp_transport}). By default, FFMPEG uses UDP for RTSP streams. It sets HLS segment duration (\mintinline{bash}{-hls_time}) to 3 seconds and the number of segments per playlist (\mintinline{bash}{-hls_wrap}) to 10. The final M3U8 file could be served in the web browser directly within the \mintinline{html}{<video>} HTML tag. The M3U8 acts as a playlist that tells the browser which segment file to playback. The segment file is an MPEG Transport Stream (MPEG-TS) file that can be decoded in the web browser.

Our evaluation of the second approach is based on the combination of MPEG and WebSockets. It uses the similar FFMPEG command with different attributes \mintinline{bash}{-f mpegts} and \mintinline{bash}{-codec:v mpeg1video} with the output relayed to the Websocket stream, which will be later decoded in the client browser (within the  \mintinline{html}{<canvas>} HTML tag). The \mintinline{bash}{-f mpegts} option specifies the output format as MPEG-TS. This format is commonly used for streaming multimedia data, especially over UDP or multicast networks. The \mintinline{bash}{-codec:v mpeg1video} sets the video codec to MPEG-1. This codec was chosen for compatibility reasons and remains a widely supported industry standard.

The third approach explores WebRTC, which allows real-time communication between web browsers and applications. It achieves this by enabling peer-to-peer connections, handling NAT and firewall traversal, and offering secure media streaming capabilities, all through standardized protocols and JavaScript APIs. We use a media server to facilitate this capability by transcoding the RTSP source and serving the client browser. The signaling mechanism is done by using Websockets, while the video is served using the \mintinline{html}{<video>} HTML tag.

We are interested in the performance of video streaming; hence, we investigated their latency and start-up delays. The results, which are the average of 10 measurements, are shown in Figure~\ref{fig:vis_res}. As the video stream reference, we use the VLC media player. We measure the start-up delay when we launch the RTSP source with the VLC media player until the video stream appears on the web browser. To measure the transmission latency, we track the timestamp difference between the video stream of the VLC media player and the video stream on the web browser. 


\begin{figure}[t]
  \centering
      \includegraphics[width=0.5\textwidth]{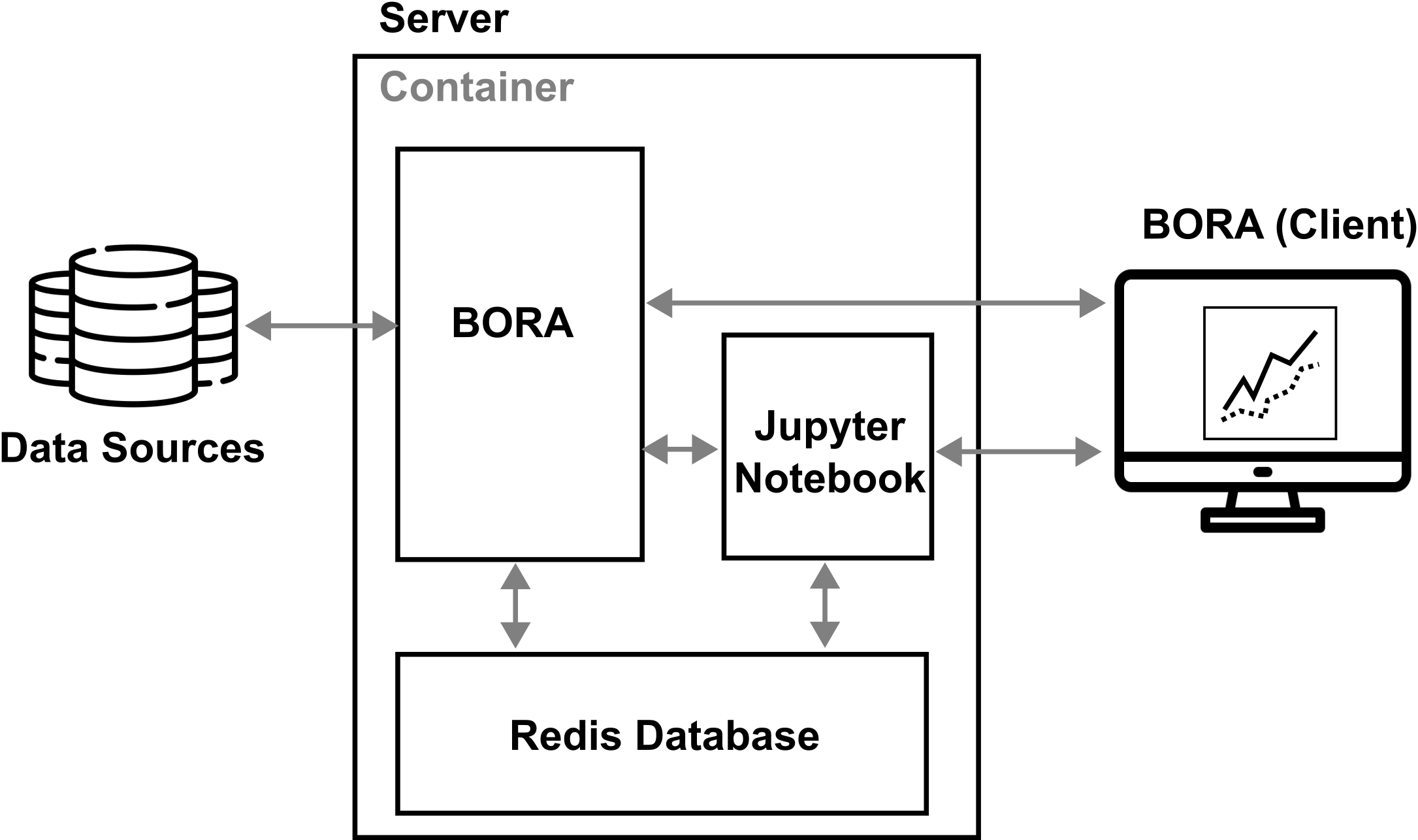}
  \caption{Data flow of the BORA framework.}
  \label{fig:bora_data_flow_general}
\end{figure}

The zero start-up delay for the HLS method should not be interpreted as a good performance indicator, mainly because this approach's transcoding latency is not captured in this performance test. During start-up, the browser serves the readily transcoded M3U8 playlist data without additional data processing. The MPEG-WS has the highest start-up delay as the video stream is transcoded upon starting the client browser. The start-up delay for the WebRTC is relatively small as the signaling mechanism to prepare for the peer-to-peer connection has a lower computation cost than the on-the-fly FFMPEG transcoding.

The transmission latency is the core performance metric related to the client browser's real-time performance. The HLS and MPEG-WS approaches have similar latency because they decode the same file format in the client browser, i.e., MPEG-TS. The WebRTC approach decodes the file format directly from the source. In our case, we use VP8 as our source format. Decoding VP8 typically involves decoding a compressed video stream using the VP8 video codec, which may require less computational resources than decoding MPEG-TS~\cite{tidestrom2019investigation}. Decoding MPEG-TS involves not only parsing the container format but also decoding the compressed video streams within it. WebRTC is the top choice in the assessment because of its minimal transmission latency and start-up delay.

\section{Implementation and Example Applications}


We implement our concepts in BORA, a browser-based system for data quality monitoring of experiment data. BORA is a lightweight framework emphasizing low development effort and supporting video streaming. The framework is based on a client-server architecture, with the server component mainly parsing the different incoming protocols. Figure~\ref{fig:bora_data_flow_general} shows the general data flow of the BORA framework, where BORA caches the sensors' recent data from a variety of sources and will be used
by Jupiter notebooks for near real-time analysis and trend
visualization or prediction. We use the Redis database to store these recent data.

\begin{figure}[t]
  \centering
      \includegraphics[width=0.38\textwidth]{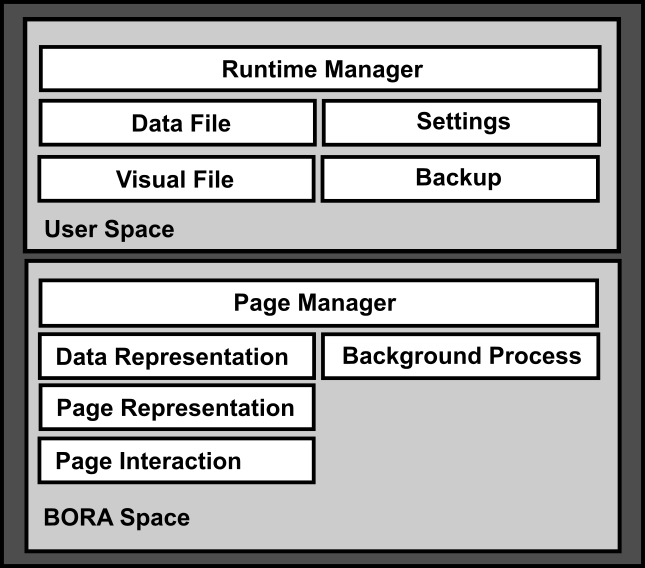}
  \caption{Components of the BORA architecture.}
  \label{fig:bora_components}
\end{figure}

\begin{figure*}[t]
  \centering
      \includegraphics[width=0.86\textwidth]{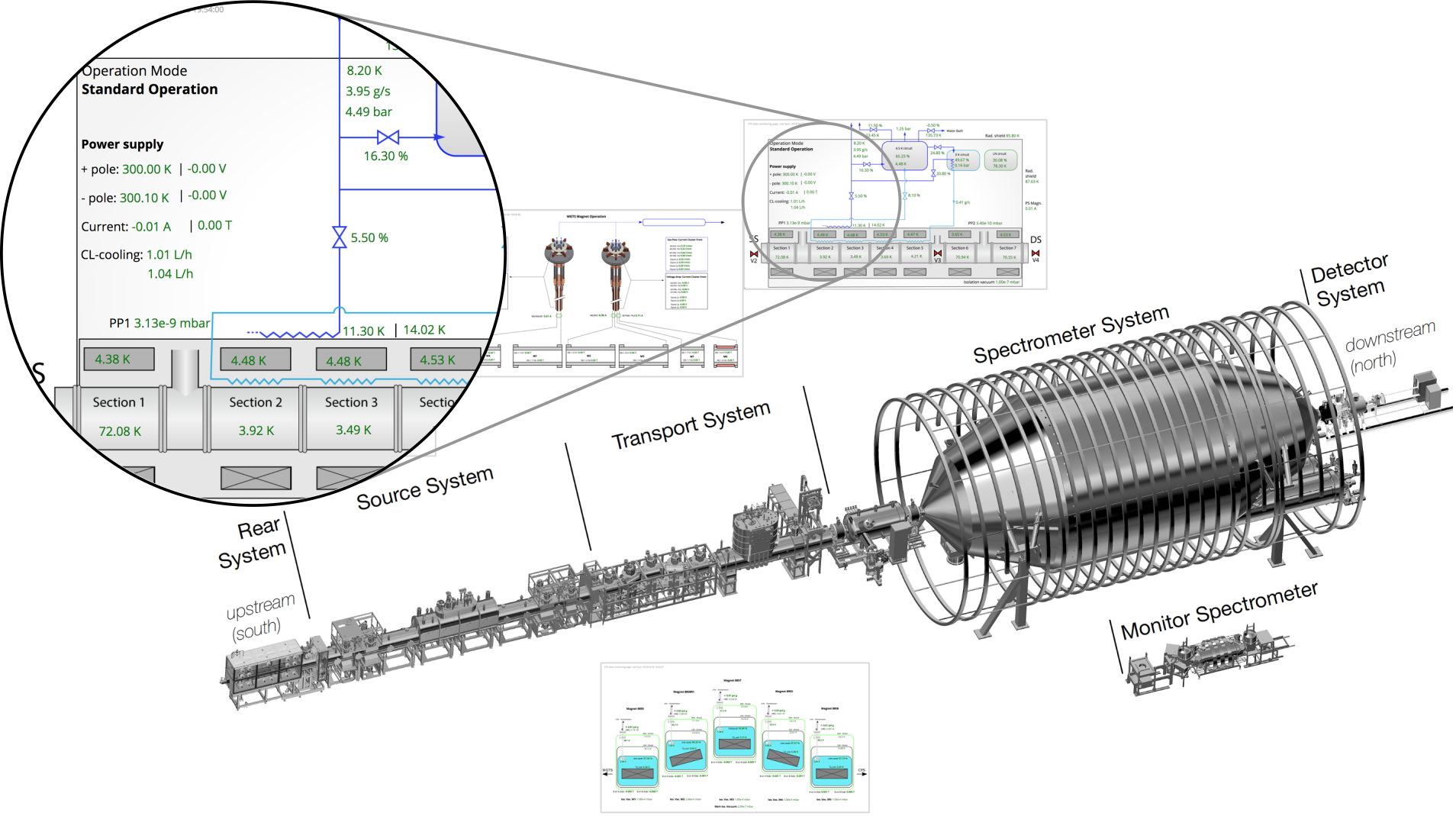}
  \caption{The KATRIN experiment comprises multiple systems. Grey boxes depict some of the BORA data displays deployed for the experiment.~\cite{aker2021design}. Currently, 22 active BORA data displays are being installed to monitor the health of the experiment.}
  \label{fig:katrin_teaser}
\end{figure*}

Figure~\ref{fig:bora_components} depicts the components of the BORA architecture, which comprises two separate design spaces: the user space and the BORA space. In the user space, the \mintinline{bash}{Runtime Manager} handles the sub-components files and backup, while in the BORA space, the \mintinline{bash}{Page Manager} handles the sub-components representation, interaction, and background processes. The BORA space operates as the central entity tasked with managing the layout representation and the logic integration for the final deployment. Within this context, we define every entity as a widget. For example, a data widget only displays data, while an input box widget can both display and manipulate data. Hence, a visual representation and interaction form the foundation of a widget. The widget's appearance is denoted by the components of \mintinline{bash}{Data Representation} and \mintinline{bash}{Page Representation}, while the logical interaction is regulated by the components of \mintinline{bash}{Page Interaction} and \mintinline{bash}{Background Process}. In terms of visual aspects, the \mintinline{bash}{Data Representation} establishes the widget attributes to be loaded on the designer page, while the \mintinline{bash}{Page Representation} governs the visual representation of each widget on the HTML page. The user space is a layer that encompasses the data display's specific information. The \mintinline{bash}{Runtime Manager} component will load the information from the settings, data file, and visual file components and then convert this information into front-end static files.

Over the course of development, we have built many data displays with BORA in real-world experiments. We now describe several examples of applications that convey the representative usage and unique capabilities of BORA. The full source code is available online at \url{https://github.com/kit-ipe/bora}.

\subsection{Status Display for the KATRIN experiment}

The KATRIN experiment~\cite{aker2021design} encompasses several systems, each responsible for handling hundreds of parameters (Figure~\ref{fig:katrin_teaser}). The operator must monitor a few specific system parameters to ensure the experiment's operational health. The parameters are not limited solely to a single system, but cross-system parameters should be accessible as well. Since the KATRIN experiment data is stored in the ADEI~\cite{chilingaryan2010advanced} system, we created BORA displays that query data from the ADEI's \mintinline{bash}{getdata.php} interface. The data is polled at regular intervals, and the data is refreshed on the final page. 

Figure~\ref{fig:katrin_teaser} also shows BORA displays with different background images, accommodating the different visual requirements of the subsystems. In this example, we use an ADEI parser in the BORA server. However, this parser could be replaced by other communication protocols, depending on the experiment. There are currently 22 active BORA status displays that report the health of the different subsystems.

\subsection{Interactive Data Display for High-speed Imaging Camera}

BORA is not limited to read-only displays. In the initial stages of BORA development, it became evident that a data display with read-only mode has significant limitations. Consequently, scientists aspire to manage the data acquisition parameters and concurrently process the data stream in real-time. To support this feature, the functionality of widgets can be enhanced by implementing AJAX calls. 

\begin{figure}[t]
  \centering
      \includegraphics[width=0.5\textwidth]{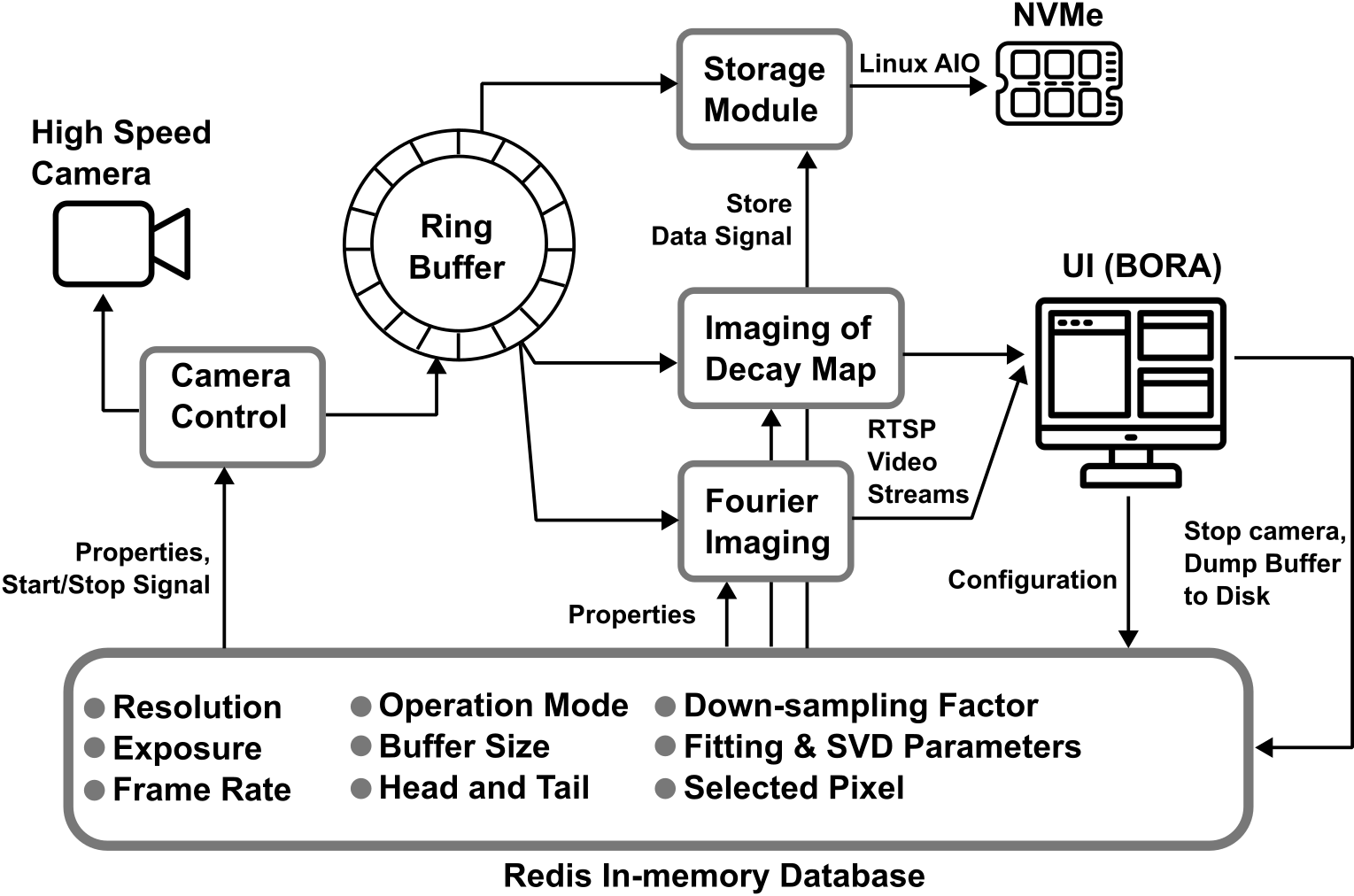}
  \caption{Data flow of the system.}
  \label{fig:data_flow}
\end{figure}

In material science, the proximity effect in excitonic materials is quickly gaining prominence as an influential tool for customizing unique quantum behaviors and functions. Yet, there is a notable absence of real-time observation and thorough mechanistic understanding of these effects at the nanoscale. Our proposed monitoring system design will rely on a small set of straightforward concepts to optimize data flow. The illustration in Figure~\ref{fig:data_flow} represents the data flow. The system allocates a significant memory region during boot-up to form a ring buffer. The camera Software Development Kit (SDK) is set to stream frames directly into this buffer. The recorded data remains stationary within the software components, as each component directly accesses the corresponding location in the ring buffer. The data processing software eliminates noise, performs online data reduction, and extracts relevant information about the studied process. The information consists of numerical values shared with other system components using a fast in-memory database (Redis~\cite{carlson2013redis}) and multiple video streams delivered to the operator monitoring experiment via the Real Time Streaming Protocol (RTSP)~\cite{schulzrinne1998real}. We convert the online monitoring into a web application to enable remote operation. The web application includes multiple views that integrate video streams, relevant parameters from the in-memory database, and a set of control set-points for adjusting all system parameters. 

\begin{figure}[t]
  \centering
      \includegraphics[width=0.5\textwidth]{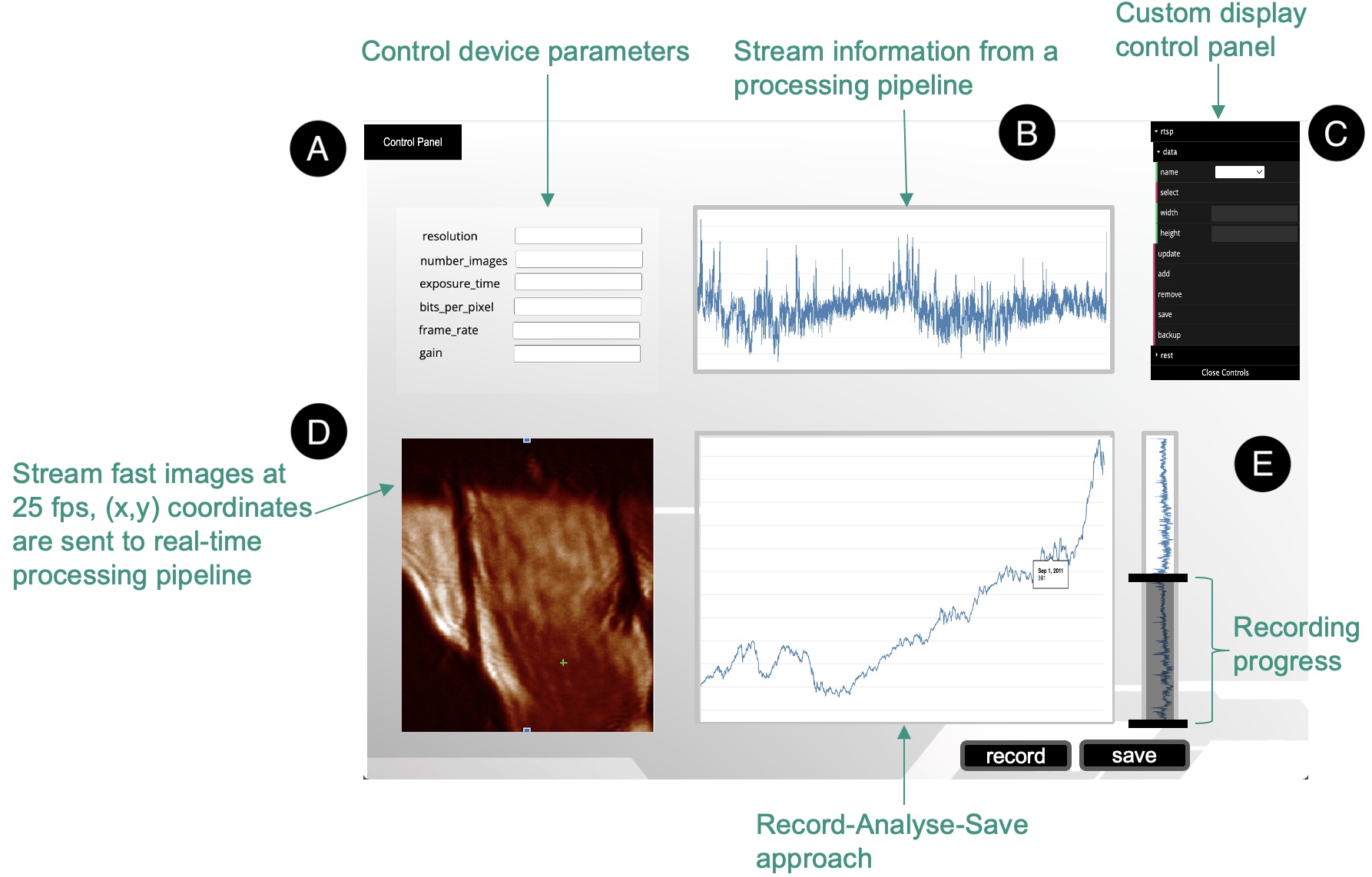}
  \caption{Interface overview: We created a BORA display showing high frame-rate data and other plugins.}
  \label{fig:bora_gui}
\end{figure}

The interface overview of a BORA data display is shown in Figure~\ref{fig:bora_gui}. In this image, a display showcases the high frame-rate data and other plugins. Figure~\ref{fig:bora_gui}-A contains the plugin that sends data to the server for updating device parameters. Figure~\ref{fig:bora_gui}-B shows a time-series graph of a parameter given by the server's processing pipeline. Figure~\ref{fig:bora_gui}-C on the selection panel allows users to select the plugin they want. Figure~\ref{fig:bora_gui}-D represents the video stream generated by processing the high frame-rate images. Figure~\ref{fig:bora_gui}-E provides video stream information and permits the selection of video segments to be recorded.


\subsection{Status Monitoring of Embedded System Applications}

Until this point, we have implicitly assumed that the BORA software stack is deployed on full-scale data centers and PC systems. However, fully autonomous measurement systems in the form of embedded systems and Systems on Chip (SoC) have grown in relevance for many scientific applications.

Such examples include the DTS-100G DAQ platform, used for cryogenic sensor readout~\cite{dts100}, or the HighFlex2 DAQ platform as shown in~\cite{highflex}. Both systems feature a modern Zynq MPSoC-based system combining a powerful FPGA and ARM-based processor. Thanks to this combination, these platforms can operate independently of a host computer. The presence of an ARM processor on those systems allows the deployment of (mostly) regular operating systems capable of running conventional software applications. 

Due to their standalone nature, systems like these also require independent monitoring and control capabilities. We envision BORA as a self-contained integrated status monitoring system for such systems, in the form of a containerized application, that can be deployed on the ARM SoC of these platforms. Both the BORA server components and a small web server that serves the status monitor web view can simultaneously be deployed on the SoC in these environments. Live measurement data is accessible to these SoCs through inter-fabric communication with the FPGA (such as the AXI Protocol) and can be processed by the BORA server locally before being provided as a remote monitoring web view over the Ethernet interfaces of these SoCs. Integrating BORA into such standalone systems thus increases their feasibility as fully self-contained measurement devices without additional external software support for operation and status monitoring. The feasibility of this approach is currently under investigation.

\section{Conclusion}

In this paper, we contribute design concepts and also provide an implementation for large-scale experiments. First, we give an overview of the existing data monitoring systems developed by other experiments and summarize the similarities between these systems. We focus on simplicity and lightweight deployment while providing a scalable system that can accommodate experiments with different requirements. As our framework supports complex data encoding via video streaming, this study thoroughly examines real-time video streaming methods for monitoring high frame rates through web browsers as the client medium. The evaluation assessed the start-up delay and transmission latency of HLS, MPEG-Websocket, and WebRTC methods, concluding that WebRTC is optimal for a low-latency system. 

\bibliographystyle{ieeetr}
\bibliography{tns}

\end{document}